\documentclass{revtex4}
\usepackage{amssymb,amsmath,amsfonts}
\usepackage{graphicx,graphics}
\DeclareGraphicsExtensions{.jpg,.jpeg,.pdf,.png,.mps,.eps,.ps,.gif}
\usepackage{color}

\usepackage{framed}
\newcommand{\rem}[1]{}
\newcommand\etal{\mbox{\textit{et al.}}}
\DeclareMathAlphabet{\mathbi}{OML}{cmm}{b}{it}

\newcommand{\bx}{\mathbi{x}}

\newcommand{\bel}{\begin{equation}\label}
\newcommand{\ee}{\end{equation}}
\newcommand{\beq}{\begin{eqnarray}\label} 
\newcommand{\eeq}{\end{eqnarray}} 
\newcommand{\bc}{\begin{center}} 
\newcommand{\ec}{\end{center}} 
\newcommand{\ben}{\begin{enumerate}}
\newcommand{\een}{\end{enumerate}}
\newcommand{\bit}{\begin{itemize}}
\newcommand{\eit}{\end{itemize}}
\newcommand{\I}{\int_{\mathcal{V}}}
\newcommand{\bom}{\mbox{\boldmath$\omega$}}
\newcommand{\bdf}{\mathbi{f}}
\newcommand{\bk}{\mbox{\boldmath$k$}}
\newcommand{\bu}{\mbox{\boldmath$u$}}
\newcommand\shalf{\ensuremath{{\scriptstyle\frac{1}{2}}}}
\graphicspath{{matlab/}}
\def\la   {\left<}
\def\ra   {\right>}
\def\li   {\rightarrow}

\begin{document}
\rem{

\author{D. A. Donzis}
\affiliation{Department of Aerospace Engineering, Texas A\&M University, College Station, Texas, TX 77840, USA}
\email{donzis@tamu.edu}
\homepage{http://people.tamu.edu/~donzis}
\author{J. D. Gibbon}
\affiliation{Department of Mathematics, Imperial College London, London SW7 2AZ, UK}
\email{j.d.gibbon@ic.ac.uk}\homepage{www2.imperial.ac.uk/~jdg}
\author{A. Gupta and R. Pandit\footnote{Jawaharlal Nehru Centre for Advanced Scientific Research, Bangalore, India.}}
\affiliation{Department of Physics, Indian Institute of Science, Bangalore 560 012, India}
\email{rahul@physics.iisc.ernet.in}
\homepage{http://www.physics.iisc.ernet.in/~rahul}
\author{R. M. Kerr}
\affiliation{Department of Mathematics, University of Warwick, Coventry CV4 7AL, UK}
\email{robert.kerr@warwick.ac.uk}
\homepage{http://www.eng.warwick.ac.uk/staff/rmk}
\author{D. Vincenzi}
\affiliation{CNRS, Laboratoire Jean-Alexandre Dieudonn\'{e},  
Universit\'{e} de Nice Sophia Antipolis, Nice  06050, France}
\email{dario.vincenzi@unice.edu}
\homepage{http://math.unice.fr/~vincenzi}
}
\bc
\textbf{\large Vorticity moments in four numerical simulations\\
of the $3D$ Navier-Stokes equations}
\ec
\bc
\textbf{D. Donzis}\\
$^1$Department of Aerospace Engineering, Texas A\&M University,\break College Station, Texas, TX 77840, USA
\ec
\bc
\textbf{J. D. Gibbon}\\
Department of Mathematics, Imperial College London, London SW7 2AZ, UK
\ec
\bc
\textbf{A. Gupta}\\
Department of Physics, Indian Institute of Science, Bangalore 560 012, India
\ec
\bc
\textbf{R. M. Kerr}\\
Department of Mathematics, University of Warwick,  Coventry CV4 7AL, UK
\ec
\bc
\textbf{R. Pandit}\\
Department of Physics, Indian Institute of Science, Bangalore 560 012, India\\
and\\
Jawaharlal Nehru Centre for Advanced Scientific Research, Bangalore, India
\ec
\bc
\textbf{D. Vincenzi}\\
CNRS, Laboratoire Jean-Alexandre Dieudonn\'{e},\\
Universit\'{e} de Nice Sophia Antipolis, Nice 06050, France
\ec

\begin{abstract}
The issue of intermittency in numerical solutions of the $3D$ Navier-Stokes equations on a periodic box $[0,\,L]^{3}$ 
is addressed through four sets of numerical simulations that calculate a new set of variables defined by $D_{m}(t) = \left(\varpi_{0}^{-1}\Omega_{m}\right)^{\alpha_{m}}$  for $1 \leq m \leq \infty$ where $\alpha_{m}= \frac{2m}{4m-3}$ 
and $\left[\Omega_{m}(t)\right]^{2m} = L^{-3}\I |\bom|^{2m}dV$ with $\varpi_{0} = \nu L^{-2}$. All four simulations 
unexpectedly show that the $D_{m}$ are ordered for $m = 1\,,...,\,9$ such that $D_{m+1} < D_{m}$. Moreover, the $D_{m}$ 
squeeze together such that $D_{m+1}/D_{m}\nearrow 1$ as $m$ increases. The first simulation is of very anisotropic decaying 
turbulence\,; the second and third are of decaying isotropic turbulence from random initial conditions and forced isotropic 
turbulence at constant Grashof number respectively\,; the fourth is of very high Reynolds number forced, stationary, isotropic 
turbulence at up to resolutions of $4096^{3}$. 
\end{abstract}

\maketitle

\section{Introduction}\label{intro}

\subsection{Background}\label{back}

Intermittency in both the vorticity and  strain fields is a dominant feature of developing and  developed turbulence.  
It has been studied extensively both experimentally (Sreenivasan 1985, Meneveau and Sreenivasan 1991) and 
numerically (Kerr 1985, Jimenez, Wray, Saffman and Rogallo 1993, Donzis, Yeung and Sreenivasan 2008, Ishihara, 
Gotoh and Kaneda 2009, Donzis and Yeung 2010,  Donzis, Sreenivasan and Yeung 2012, Yeung, Donzis and 
Sreenivasan 2012). Statistical physicists generally use velocity structure functions to study this phenomenon and 
have diagnosed the degree of intermittency by how much the velocity structure function exponents $\zeta_{p}$ 
differ from linear when $p>3$ (Frisch 1995, Schumacher, Yakhot and Sreenivasan  2007,  Boffetta, Mazzino and 
Vulpiani 2008, Pandit, Perlekar and Ray 2009).  The standard way to quantify equal-time, multi-scaling exponents 
is a challenging numerical task (Arneodo \etal~2008, Ray, Mitra and Pandit 2008, Ray, Mitra, Perlekar and Pandit 
2011). The multi-scaling approach is even more challenging for the 3D Navier-Stokes equations
\bel{NS1}
\partial_{t}\bu + \bu\cdot\nabla\bu = \nu \Delta\bu -\nabla P\qquad\qquad \mbox{div}\,\bu = 0
\ee
because the velocity field $\bu(\bx,\,t)$ and pressure $P(\bx,\,t)$ evolve in time, so, in general, time-dependent 
structure functions must be used to study dynamic multi-scaling (Ray, Mitra and Pandit 2008, Ray, Mitra, Perlekar 
and Pandit 2011). This paper will introduce an analysis of some new and existing numerical computations that 
gives new insights into and provides a new method for distinguishing alternative regimes of behaviour in the vorticity 
field. To explain the nature of these regimes, let us consider the vorticity field $\bom = \mbox{curl}\,\bu$ on a finite 
periodic domain $[0,\,L]^{3}$ within the setting of the volume integrals which define a set of frequencies 
\bel{Omdef}
\Omega_{m}(t) = \left(L^{-3}\I |\bom|^{2m}dV\right)^{1/2m}\,,\qquad\qquad 1 \leq m \leq \infty\,.
\ee
Some recent work has centred around a dimensionless re-scaling of the $\Omega_{m}$ such that 
(Gibbon 2010,\,2011,\,2012a,b)
\bel{Dmdef}
D_{m}(t) = \left(\varpi_{0}^{-1}\Omega_{m}\right)^{\alpha_{m}}\,,\qquad\qquad\alpha_{m} = \frac{2m}{4m-3}\,,
\ee
where $\varpi_{0}$ is a fixed frequency defined by $\varpi_{0} = \nu L^{-2}$.  The origin of this re-scaling, valid for 
both the Navier-Stokes and Euler equations, has been explained elsewhere (Gibbon 2011,\,2012a,b) where it has been 
shown that, with additive $L^2$-forcing, weak solutions obey the time average up to time $T$
\bel{Dmav}
\left<D_{m}\right>_{time~av.}
\leq c\,Re^{3} + O\left(T^{-1}\right)\,.
\ee
The first in the hierarchy, $D_{1} = \varpi_{0}^{-2}Z$, is proportional to the global enstrophy $Z = \Omega_{1}^{2}$
and may be insensitive to deep fine-scale fluctuations. The higher $D_{m}$ may be more sensitive so their measurement 
over a wide range of $m$ could be a useful diagnostic of intermittency. However, the end of the sequence, $D_{\infty}(t)$, 
is hard to measure numerically, especially in highly intermittent flows. While H\"older's inequality enforces a natural ordering 
on the frequencies $\Omega_{m}$ such that $\Omega_{m} \leq \Omega_{m+1}$  for $1 \leq m \leq \infty$, no such 
natural ordering is enforced on the $D_{m}$ because the $\alpha_{m}$ \textit{decrease} with $m$. Thus there are two 
possible regimes\,:
\bel{regI}
D_{m+1}(t) < D_{m}(t)\,,\quad\mbox{(regime~I)}\,,\qquad\qquad
D_{m}(t) \leq D_{m+1}(t)\,,\quad\mbox{(regime~II)}\,.
\ee
The issues to be addressed in this paper in our four numerical simulations of the $3D$ Navier-Stokes equations are\,: 
\ben\itemsep 0mm
\item Which of these regimes is favoured or is there an oscillation between them?  If one regime is favoured, 
are the $D_{m}$ well separated? What is the role of the enstrophy $D_{1}$?
\item Are these processes $m$-dependent?
\item Are they $Re$-dependent?
\item Are they dependent upon initial conditions?
\een

\subsection{Simulations used for tests}\label{sim}

An important point with respect to numerical simulations of the scaled higher order moments $D_{m}(t)$ is 
that their ratios might converge better than their actual values.  This is consistent with the results reported in 
Yeung \etal~(2012) and Donzis \etal~(2012) where convergence for the 
ratios of higher-order vorticity and dissipation (strain) moments were obtained, even when the statistics of the 
individual moments showed no evidence of convergence (Kerr 2012a).  This answered a problem first raised 
in Kerr (1985) where it was noted that in forced simulations at modestly high Reynolds numbers, the averages 
of the vorticity and strain moments above 6-th order did not converge.  The determination of the $D_{m}(t)$ in 
simulations is not difficult whereas, in contrast, traditional numerical tools such as higher-order structure 
functions require a combination of larger domains and finer resolution than is currently feasible. 
This paper will calculate and compare the $D_{m}(t)$ from four data sets\,: two where the
average kinetic energy $E=L^{-3}\int_{\cal V}\shalf|\bu|^{2}\,dV$ decays in time, and two where $E$ is held 
approximately constant by forcing at the low wavenumbers. The first is a unique data set from a computation 
in which fully-developed turbulence forms from the interaction of two anti-parallel vortices and whose kinetic 
energy $E$ decays strongly after the first peak in the normalised enstrophy production $-S_u$.
Because this calculation has not been fully described 
before, some introductory discussion is provided at the start of that section (\S\ref{RMK}). The other three 
data sets represent more traditional decaying and forced homogeneous, isotropic numerical turbulence. In the 
decaying calculations in \S\ref{RMK} and the decaying and forced calculations in \S\ref{RP} the moments have 
been determined relatively continuously in time which makes a helpful comparison with the results of \S\ref{intro}.  
For the fourth data ($4096^3$) set of \S\ref{DD} (Yeung \etal~2012, Donzis \etal~2012), a similar conclusion is 
reached by studying the dependence of the average value of $D_{m}$ on the Reynolds number.
 
An advantage of the first data set described in \S\ref{RMK} is that the predicted convergence properties of 
ratios of the $D_m(t)$ can be tested for a calculation with huge fluctuations in the production of enstrophy, 
and therefore in the higher $D_{m}(t)$.  That the calculation eventually exhibits traditional turbulent statistics 
and spectra is a bonus in justifying its use.  However, this new initial condition is very specialized and any trends 
need to be confirmed using a more traditional decaying homogeneous, isotropic data set, which is the purpose 
of the second calculation discussed in \S\ref{RP}.  \S\ref{RP} also contains forced simulations of homogeneous 
and isotropic turbulence at constant Grashof number. Finally, the fourth calculation in \S\ref{DD} provides validation 
from a forced, massively parallel, pseudo-spectral calculation ($4096^{3}$ with $R_{\lambda} \approx 1000$) 
calculation to show that these trends are not restricted to low or moderate Reynolds numbers. 
Assessing the scaling of moments of intermittent quantities such as vorticity, strain rates or velocity gradients 
has been a critical component of characterizing and understanding intermittency. Of particular interest is how 
these moments scale with the Reynolds number, which is typically high in applications. At the same time, 
different orders provide information about fluctuations of different intensities. Low and high-order moments, 
for example, are associated with weak and strong fluctuations. Thus, the understanding of the scaling of the 
moments $D_{m}$, especially at high $m$, can also shed light on the nature of intermittency and the most 
extreme events in turbulence.

\subsection{A summary of results}\label{sum}

The simulations described in \S\ref{RMK}, \S\ref{RP} and \S\ref{DD}, and illustrated in  Figs. \ref{Kerr}, 
\ref{Bang} and \ref{TAMU_rat}, each observe that a strict ordering of the $D_{m}$ occurs, as in regime I\,; 
namely $D_{m+1} < D_{m}$ (on log-linear plots). To assess the significance of this, we write down the 
relation $D_{m+1} < D_{m}$ in terms of $\Omega_{m}$ and use H\"older's inequality $\Omega_{m} 
\leq \Omega_{m+1}$ on the extreme left hand side 
\bel{3sim1A}
\varpi_{0}^{-1}\Omega_{m} \leq \varpi_{0}^{-1}\Omega_{m+1} < 
\left(\varpi_{0}^{-1}\Omega_{m}\right)^{\alpha_{m}/\alpha_{m+1}}\,.
\ee
As $m\to\infty$, $\alpha_{m}\searrow\alpha_{m+1}$, and so (\ref{3sim1A}) shows that $\Omega_{m+1}/
\Omega_{m} \searrow 1$. Thus, in regime I the $\Omega_{m}$ must be squeezed 
together for high $m$.  In terms of the $D_{m}$  (\ref{3sim1A}) is written as
\bel{3sim2}
D_{m}^{\alpha_{m+1}/\alpha_{m}} \leq D_{m+1} < D_{m}\,.
\ee
While respecting the ordering $D_{m+1} < D_{m}$ , $D_{m+1}$ is squeezed up close to $D_{m}$ as $m\to\infty$
\bel{3sim3}
\lim_{m\to\infty}\frac{D_{m+1}}{D_{m}}\nearrow 1\,.
\ee
This squeezing phemonenon is observed in all four data sets where the $D_{m}$-curves lie very close for $m>3$ as in 
Figs. \ref{Kerr}, \ref{Bang} and \ref{TAMU_rat}. Moreover, the values of $D_{1}$ in all four simulations lie far above the 
rest of the $D_{m}$ giving rise to a suggestion, explored in \S\ref{con}, that a depletion of nonlinearity is occurring which 
could be the cause of Navier-Stokes regularity. The most extreme intermittent events are represented by moments at 
increasingly large $m$. Our results suggest the saturation of these high order moments. This is significant as it constrains 
the shape of the tails of the PDF of vorticity which has been the focus of intense investigations (Kerr 1985, Jimenez \etal~1993, 
Donzis \etal~2008,  Ishihara \etal~2009, Donzis and Yeung 2010, Yeung \etal~2012, Donzis \etal~2012). The fourth data set 
(forced, stationary, isotropic turbulence), the results of which are displayed in \S\ref{DD}, furnishes us with the opportunity to 
compare these results with other results on intermittency available in the literature. For example, within the multifractal model, 
Nelkin (1990) found that normalized moments of velocity gradients scale as
\bel{as1A}
\la u_x^p\ra/\la u_x^2\ra^{p/2}\sim {Re_\lambda}^{d_p}\,,
\ee
where $d_p$ is obtained from the multifractal spectrum and $\la \cdot\ra$ is the usual notation for the statistical average. 
Using the well-known result $\la u_x^2\ra \sim (U_{0}/L)^{2}Re_{\lambda}^{2}$ due to the dissipative anomaly, it is 
readily shown that $\la u_{x}^{p}\ra \sim {Re_\lambda}^{p+d_{p}}$. Our interest lies in the limit $p\li \infty$ where 
it can be shown that $\lim_{p\li \infty}d_p/p = c$. The constant $c$ is given by $c = 3(1-\mathcal{D}_\infty)
/(3+\mathcal{D}_\infty)$ with $\mathcal{D}_\infty$ representing the limit $\lim_{q\li \infty}\mathcal{D}_q$ of the 
generalized dimensions $\mathcal{D}_q$ (Nelkin 1990, Hentschel and Procaccia 1983).  Clearly moments of the form  
$\la u_x^p\ra^{1/p}$ saturate at high $p$, consistent with (\ref{3sim3}). While experimentally it is difficult to measure 
$\mathcal{D}_\infty$ reliably, its value appears to be smaller than 1.0 (Meneveau and Sreenivasan 1991). The ratio of 
successive orders is
\bel{as2}
\la u_x^{p+1}\ra^{1/(p+1)}/\la u_x^p\ra^{1/p}
\sim 
Re_\lambda^{(1+d_p/p)-(1+d_{p+1}/(p+1))}\,.
\ee
The limiting behavior of $d_p$ shows that $\lim_{p\li \infty}[(1+d_p/p)-(1+d_{p+1}/(p+1))]= 0$, and therefore the ratio 
on the left hand side of (\ref{as2}) tends to a constant independent of $p$ and $Re_{\lambda}$. This is consistent with 
the squeezing together of the $\Omega_{m}$ and $D_{m}$.

\section{The first set of simulations}\label{RMK}

The new vortex reconnection calculation displayed in this section addresses the following long-standing numerical 
question\,: Can an initial condition with only a few vortices generate and sustain fully-developed turbulence through 
reconnection events in a manner similar to how turbulence forms in aircraft wakes or when anti-parallel quantum 
vortex lines reconnect numerically (Kerr 2011)?  Because this initial condition has not been fully explained before, 
some of its unique features are now described. The three directions in the flow are\,: 
(i) each initial vortex primarily points in the $\pm y$-direction\,; 
(ii) separation between the vortices lies in the $z$-direction\,; 
(iii) they propagate in the $x$-direction. 
Due to the anisotropy of the flow, an anisotropic mesh and 
domain are used with a $L_x\times L_y\times L_z=2\pi(2\times8\times1)$ domain and a
$n_x\times n_y\times n_z=512\times2048\times512$ mesh, plus symmetries used in
the $y$ and $z$ directions.
\begin{figure}
\includegraphics[scale=.10]{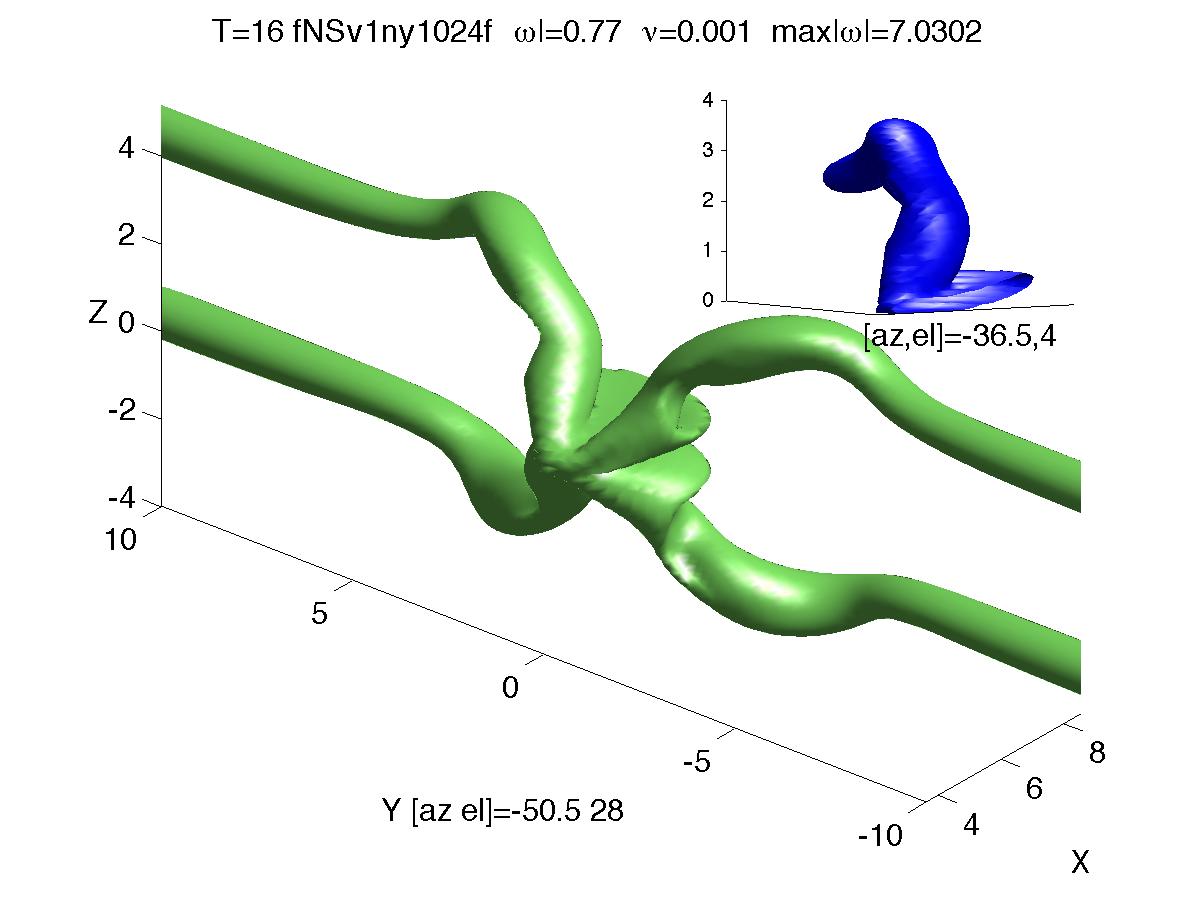}
\includegraphics[scale=.10]{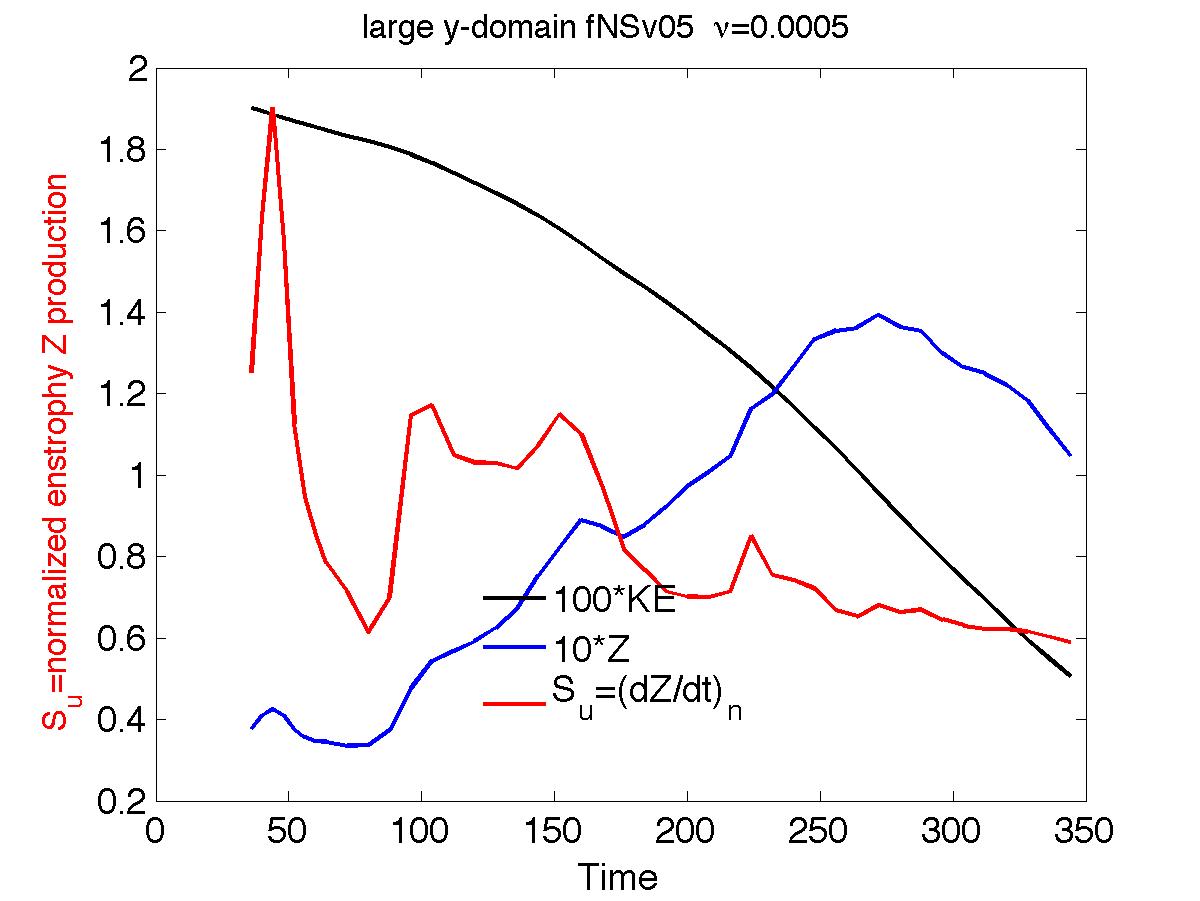}
\includegraphics[scale=.10]{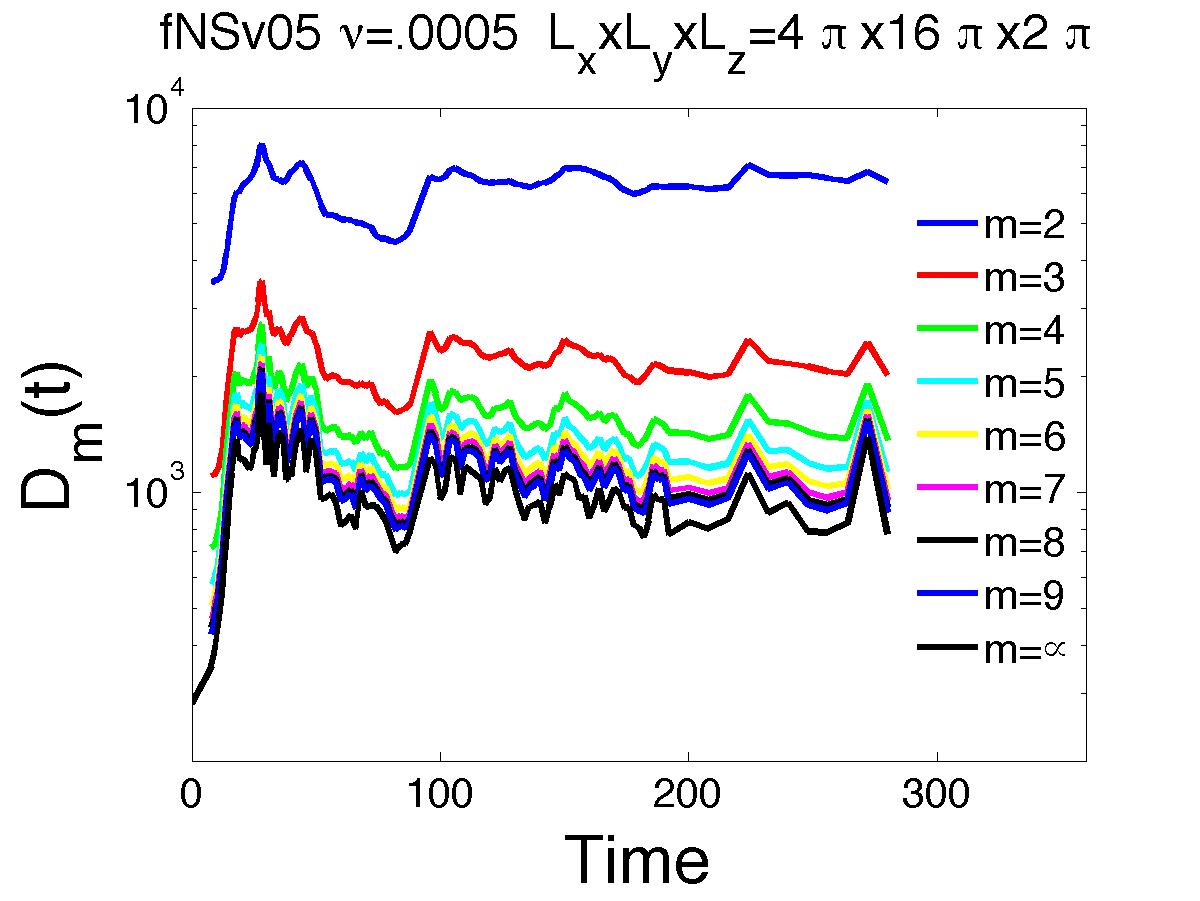} 
\caption{\scriptsize \textbf{First panel\,:} The initial condition is characterized 
by long anti-parallel vortices with a localized perturbation for the $Re=4000$ 
reconnection calculation. 
\textbf{Second panel\,:} Plots of the kinetic energy decay $E$, the enstrophy $Z$ and
its production, normalised to be consistent with experimental measurements
of the velocity derivative skewness $-S_u$. $Z$ grows until $t\approx270$, while 
$E$ is always decaying. 
\textbf{Third panel\,:} Curves range from $m=2$  to $m=9$ and include the maximum 
of vorticity $D_{\infty}$. The normalized enstrophy, $D_1$, is far above the log-scale given 
here, so it is omitted. The $D_{m}$ are ordered for all values of $Re$ and all times.}
\label{Kerr}
\end{figure}
The quantum vortex work in Kerr (2011) has shown that the two most important properties 
are that their initial perturbations need to be localized far from the periodic boundaries and 
their initial profile and direction should be balanced so that they are neither internally 
unstable nor prone to the shedding of waves or vortex sheets. The method grew from 
addressing calculations identified by Bustamante and Kerr (2008), where a single sign of 
the initial local vorticity had not been imposed rigorously. To accomplish these goals, the 
following four changes have been made to the initial condition in Bustamante and Kerr 
(2008)\,: i) The perturbation is 
strongly localised near the symmetry plane using the trajectory given in Kerr (2011)\,; ii) The 
initial vorticity profile uses the solution of a two-dimensional vortex with a smoothed core\,; 
iii) The direction of vorticity chosen at grid points follows the path of the nearest point in $3D$ 
space on the prescribed trajectory of the central vortex line\,; iv) The vortices need to be twice 
as long as in any previous anti-parallel study. Fig. \ref{Kerr} (left) shows the vorticity as the 
initial instability saturates.  By using these choices, regions of negative vorticity and vortex 
sheets on the $y=0$ plane, as described by Bustamante and Kerr (2008), are eliminated.  

The ultimate goal is to generate turbulence with a persistent -5/3 energy spectrum
and additional turbulent statistics, including the experimental velocity derivative 
skewness $S_{u} = \left<u_{x}^{3}\right>/\left<u_{x}^{2}\right>^{3/2}$, which 
is equivalent to the numerical normalized enstrophy production. The latest infinite 
$Re$-estimates of $S_u$ from forced turbulence calculations (Ishihara \etal~2009) 
find $-S_u\approx 0.68$, consistent with experimental values of $-S_u\sim 0.5-0.7$. 
Early numerical calculations showed that the $S_u$ tended to overshoot the early 
experimental values of $-S_u\approx 0.4-0.5$ before settling to the expected value 
(Orszag and Patterson 1972).  The second panel of Fig. \ref{Kerr} confirms this 
trend for the anti-parallel calculation with $-S_u$ first rising to $-S_u=1.9$ at 
$t\approx45$, then falling abruptly to $-S_u\approx0.6$ at $t\approx80$, continuing 
to fluctuate strongly between 0.6 and 1.2 for $100<t<250$, and finally decaying to 
$-S_u\approx0.6$. The full details, plus the relationship between the variations in 
$D_m$ in Fig. \ref{Kerr} and the development of swirling, turbulent vortex rings, is 
the topic of another paper (Kerr 2012b).

In Fig. \ref{Kerr}, note that for all times, all of the lower order $D_m$ ($m=1,\,\ldots,
\,9$) bound each higher-order $D_m$ (on a log-scale) which can be expressed as 
$D_{m+1}(t)<D_{m}(t)$, thus favouring regime I as in (\ref{regI}). The enstrophy 
$D_1$ lies far above all of the other curves and has been omitted. Next note a strong 
increase in the growth of each of the $D_m$, including $D_\infty$, up until $t\approx16$.  
This is the period when this calculation has nearly Euler dynamics, where the effects 
of viscosity compared to nonlinear growth are minimal. The growth of the $D_m(t)$ 
in true Euler dynamics is the topic of another paper (Kerr 2012c).

\section{The second and third set of simulations\,: DNS results for homogeneous, isotropic turbulence}\label{RP}

Data from two direct numerical simulations (DNSs) of homogeneous, isotropic 3D Navier-Stokes turbulence is now 
presented. Both of these simulations use a pseudospectral method, a $2/3$-rule for de-aliasing, and $512^3$ 
collocation points on a $[0,\,2\pi]^{3}$ domain. 
\begin{figure}
\begin{minipage}[t]{0.3\linewidth}
\includegraphics[width=\linewidth]{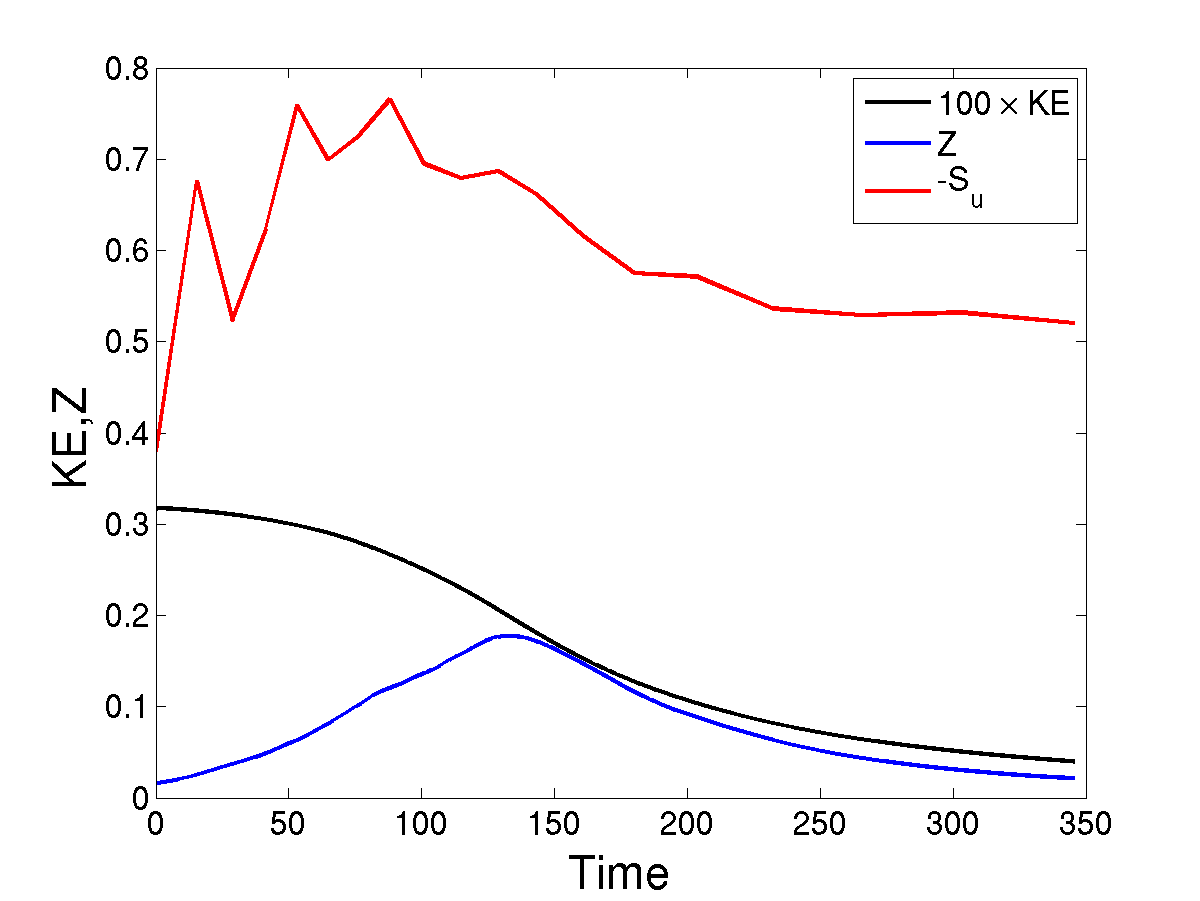}
\end{minipage} \hfill
\begin{minipage}[t]{0.3\linewidth}
\includegraphics[width=\linewidth]{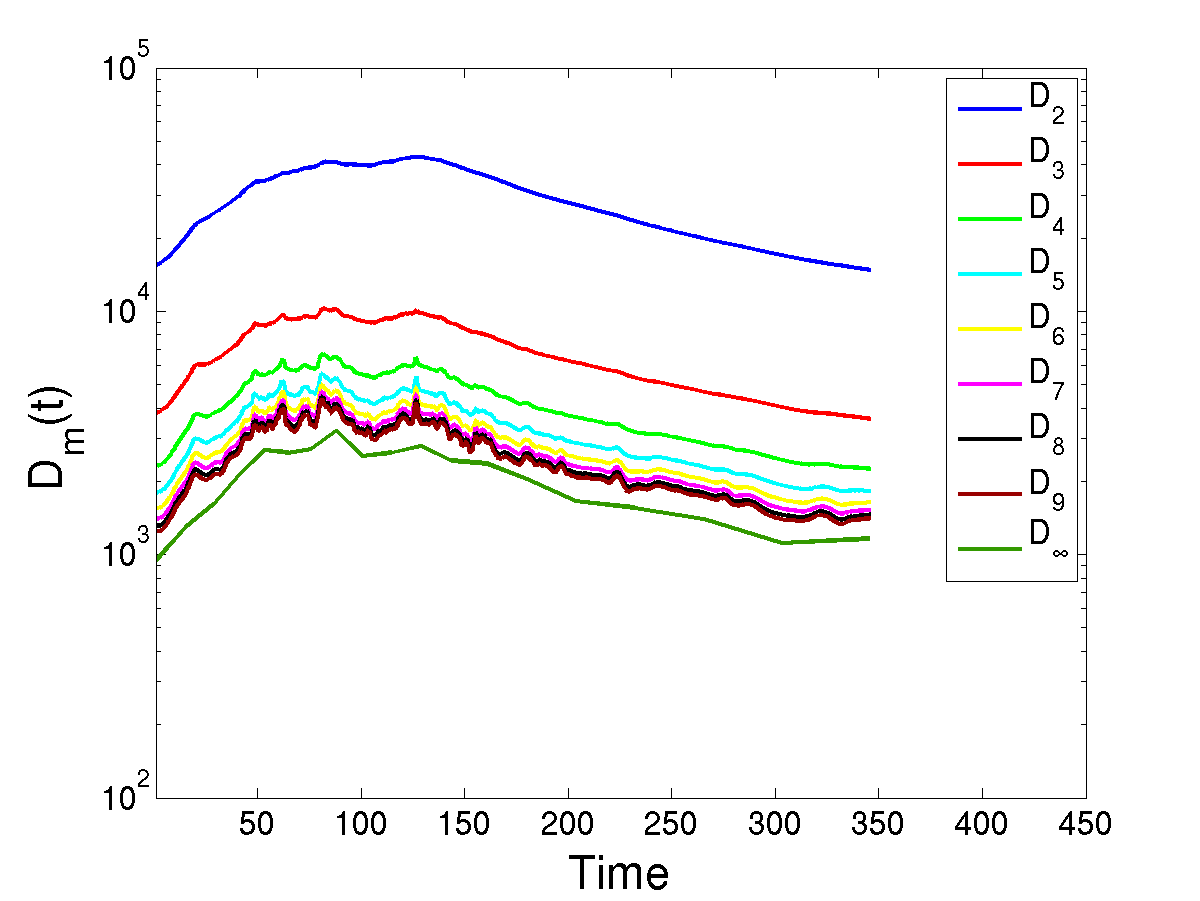}
\end{minipage} \hfill
\begin{minipage}[t]{0.3\linewidth}
\includegraphics[width=\linewidth]{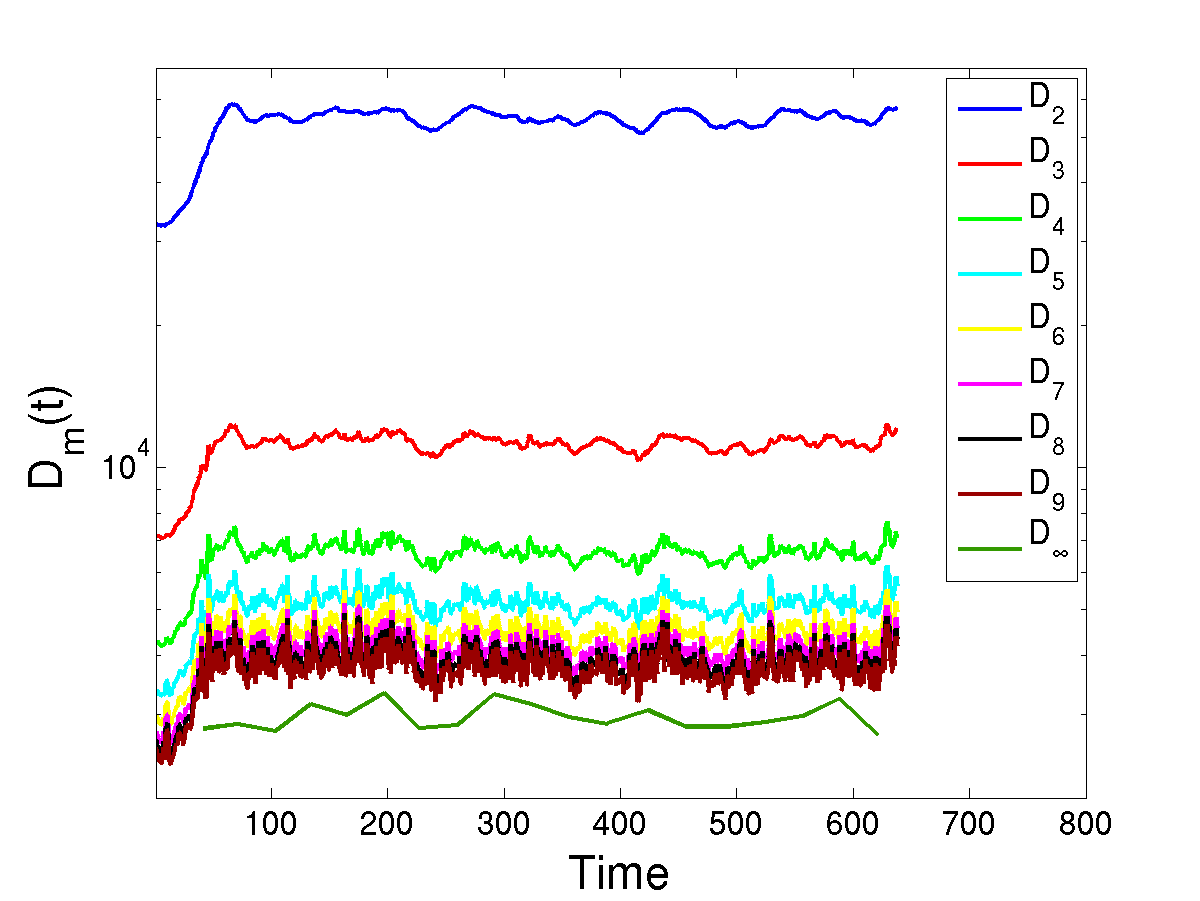}
\end{minipage} 
\caption{\scriptsize Plots versus time $t$ of the total kinetic energy (first panel, black curve), the enstrophy $Z$ (first 
panel, blue curve), the normalized enstrophy-production rate $-S_u$ (first panel, red curve), $D_m$ for $2 \leq m 
\leq 9$ (second panel, blue to brown curves), and $D_{\infty}$ (second panel, dark green curve) for our DNS of decaying, 
3D Navier-Stokes turbulence\,; the value of $D_1$ is very high, so it is omitted. The third panel is of 
statistically steady forced turbulence at constant Grashof number. 
The mean values of $D_m$ in the statistically steady state are as follows: $\left<D_1\right> = 3.1 \times 10^{11}$, 
$\left<D_2\right> = 5.5 \times 10^{4}$, $\left<D_3\right> = 1.1 \times 10^{4}$,
$\left<D_4\right> = 6.6 \times 10^{3}$,  $\left<D_5\right> = 5.2 \times 10^{3}$, 
$\left<D_6\right> = 4.5 \times 10^{3}$, $\left<D_7\right> = 4.1 \times 10^{3}$,  
$\left<D_8\right> = 3.9 \times 10^{3}$, $\left<D_9\right> = 3.7 \times 10^{3}$,
and $\left<D_{\infty}\right> = 3.0 \times 10^{3}$. Zooming in to the right panel 
makes it clear that $D_{m+1} < D_{m}$ for all values of $m$ considered.}\label{Bang}
\end{figure}
The first DNS is of decaying turbulence which reaches a Taylor-microscale Reynolds number $Re_{\lambda} \simeq 134$ 
at the main peak of the enstrophy $Z$ associated with the formation of the inertial sub-range. The Taylor-microscale 
$\lambda$ is defined in the usual way in terms of the energy spectrum $E(k)$. The initial Fourier components of the 
velocity $\tilde{\bu}_{0}(\bk)$ for the wave-vector $ k = |\bk|$ are generated by applying random phases to 
the energy spectrum $E_{0}(k) = E_{0}k^{4}\exp\{-2k^{2}\}$. 
\rem{
\bel{TMSdef}\lambda^{-1} = \left[ \int k^{2}E(k) dk / \int E(k) dk\right]^{1/2}\,.\ee
The energy spectrum and the wave-number are, respectively, $E(k)$ and $k$.  
An initial condition (subscript $0$) has been used in which the Fourier components of the velocity $\tilde{\bf{u}}_{0}(\bf{k})$
have random phases with an initial energy spectrum is $E_{0}(k) = E_{0}k^{4}\exp\{-2k^{2}\}$. 
}
The second DNS is a study of statistically steady turbulence which attains $Re_{\lambda} \simeq 182$\,; the forcing term 
$\bdf_{u}(\bx,t)$ is specified most simply in terms of $\tilde{\bdf}_{u}(\bk, t)$ whose spatial Fourier components are\,:
\bel{FC1}
\tilde\bdf_u(\bk, t) = \frac {{\cal P} \Theta (k_{f} - k)}{\sqrt {2 E_u(k_{f},\,t)}} \bu(\bk,\,t)\,,
\qquad\qquad E_{u}(k_{f},\,t) = \sum_{k\leq k_{f}}E_{u}(\bk,\,t)\,,
\ee
where $\Theta$ is the Heaviside function and $k_{f} = 2$ is the wave number above which Fourier modes are not forced. This 
forcing term maintains a constant Grashof number $Gr = L^{3}\mathcal{P}/\nu^{2} = 4.9\times10^{7}$\,: for a similar 
forcing term that holds the energy injection fixed see Sahoo, Perlekar and Pandit (2011).  

For the decaying DNS, a small inertial subrange forms at $t=100$ when the enstrophy $Z$ reaches its main peak. Assuming 
$E_u(k)=K_{0}(k)\epsilon k^{-5/3}$, the pre-factor $K_{0}(k)$ is roughly 1.5 for about half a decade of wavenumbers. 
Similar to Fig. \ref{Kerr} (second panel), Fig. \ref{Bang} (first panel) shows the time-dependence of the kinetic energy $E$, 
enstrophy $Z$ and its skewness $-S_u$.  The second and third panels in Fig. \ref{Bang} show $D_m$ versus time $t$ for 
$m=2, \dots ,9$ and $D_{\infty} = \left(\varpi_{0}^{-1}\|\omega\|_{\infty}\right)^{\alpha_\infty}$ with $\alpha_{\infty} = 
\shalf$ for both the decaying and forced DNS calculations respectively. The second and third panels also show that $D_{m} 
< D_{m+1}$ and thus demonstrate the generality of Fig. \ref{Kerr} of \S\ref{RMK}.

\rem{
The left panel of Fig. \ref{Bang} shows the time series of the kinetic energy, the enstrophy and its normalized production 
rate for the first decaying-turbulence DNS. At the main peak of $Z$ a small inertial range with a spectral scaling is obtained 
that is within error bars of the Kolmogorov spectrum $E_{u}(k) \sim k^{-5/3}$. The same spectral scaling is also obtained 
in the statistically steady state of our second DNS. The second and third panels of Fig. \ref{Bang} show plots of $D_m$ versus 
time $t$ for $m=2, \dots ,9$ and $D_{\infty} = \left(\varpi_{0}^{-1}\|\omega\|_{\infty}\right)^{1/2}$ for both the decaying 
and forced DNS calculations respectively. These make it clear that $D_{m} < D_{m+1}$ (on a log-scale) consistent with Fig. 
\ref{Kerr} of \S\ref{RMK}.}

\section{The fourth set of simulations\,: forced stationary isotropic turbulence}\label{DD}

\begin{figure}
\begin{minipage}[t]{0.4\linewidth}
\begin{tabular}{cccc}
$N$ & $Re_\lambda$ & $k_{max}\eta$ & Forcing \\ \hline 
256 & 140 & 1.4 & EP \\ 
256 & 140 & 1.4 & FEK \\ 
512 & 140 & 2.7 & FEK \\ 
2048 & 140 & 11.2 & FEK \\ 
512 & 240 & 1.4 & FEK \\ 
2048 & 240 & 5.1 & FEK \\ 
1024 & 400 & 1.4 & FEK \\ 
2048 & 400 & 2.8 & EP \\ 
2048 & 650 & 1.4 & EP \\ 
4096 & 650 & 2.7 & FEK \\ 
4096 & 1000 & 1.3 & FEK
\end{tabular}
\caption{\scriptsize Parameters of statistically stationary forced simulations\,: 
included are the resolution $N$, $Re_\lambda$,
the resolution parameter $k_{max}\eta$ and the forcing type (see text).}
\label{tab:dns_diego}
\end{minipage}
\hspace{3mm}
\begin{minipage}[t]{0.45\linewidth}
\includegraphics[width=0.6\textwidth]{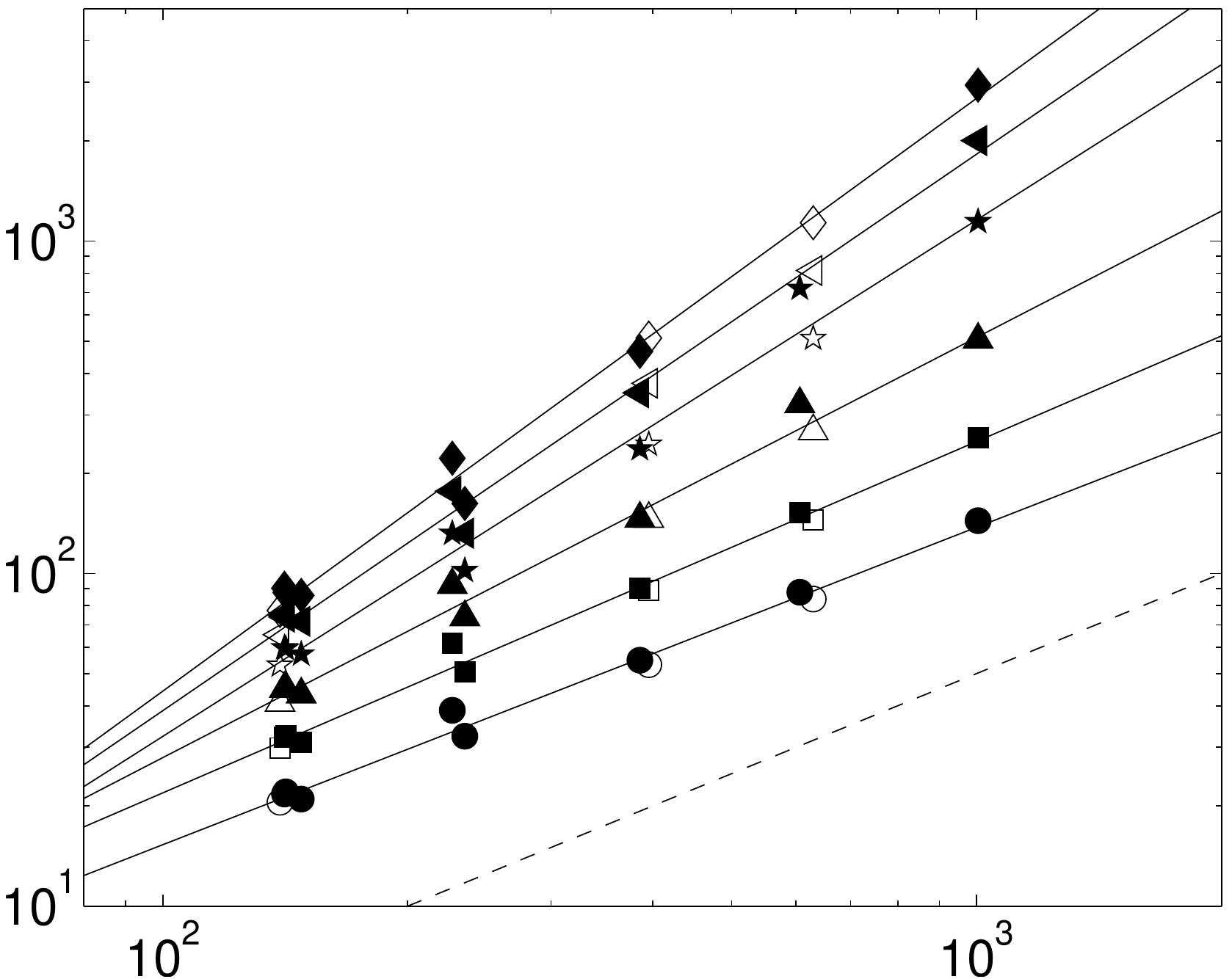}
\begin{picture}(0,0)
\put(-55,25){\vector(0,1){50}}
\put(-65,75){\scriptsize$m$}
\put(-112,75){\scriptsize$\Omega_m$}
\put(-70,-5){\scriptsize$Re_\lambda$}
\end{picture}
\caption{\scriptsize Scaling of the $\Omega_m$ as a function of 
$Re_\lambda$ for forced stationary isotropic turbulence with resolutions up to $4096^3$. 
Lines are for $m=1$ (circles), 2 (squares), 3 (triangles), 4 (stars), 
5 (left triangles), 6 (diamonds). Open and closed symbols correspond to EP and FEK 
forcing respectively. Dashed line is $\sim Re_\lambda^6$ (see text). Note that for 
$Re_\lambda\approx 650$ at $4096^3$ with FEK forcing, moments up to fourth 
order (instead of sixth) are available from our database.}\label{TAMU_om}
\end{minipage}
\end{figure}

The DNS data in this fourth set of simulations were obtained using a massively parallel pseudo-spectral 
code which achieves excellent performance on $O(10^5)$ processors. The basic numerical scheme is that 
of Rogallo (1981). The time stepping is second-order Runge-Kutta and the viscous term is exactly treated 
via an integrating factor. Aliasing errors are carefully controlled by a combination of truncation and phase 
shifting techniques. The database includes simulations with resolutions up to $4096^3$ and Taylor-Reynolds 
number up to $Re_\lambda\approx 1000$ (Donzis \etal~2012, Yeung \etal~2012). In order to maintain a 
stationary state, turbulence is forced numerically at the large scales. Since our objective is to assess the 
generality of the ordering of the moments $D_m$, here, we show results using the stochastic forcing of 
Eswaran and Pope (1988) -- denoted as EP -- as well as a deterministic scheme described in Donzis and 
Yeung (2010) -- denoted as FEK. In essence, this keeps the energy in the lowest wavenumbers fixed. For 
these two forcing schemes, the wavenumbers affected by forcing are confined to within a sphere $k < k_F$, 
where $k_F$ is of order 2 or 3.  In order to capture intense events, which are the main contributors to high-order 
moments, resolution issues have to be properly addressed. Motivated by the theoretical work of Yakhot and 
Sreenivasan (2004), resolution effects have been studied in Donzis~(2012), and Yeung \etal~(2012) with the 
conclusion that although high-order moments may be under-predicted using the standard resolution criterion -- typically 
in simulations aimed at pushing up the Reynolds number -- {\it ratios} of high-order moments are weakly 
affected by resolution issues. Small-scale resolution for a spectral simulation is typically quantified with the 
paramater $k_{max}\eta$ where $k_{max} = \sqrt{2}N/3$ is the highest resolvable wavenumber in a domain 
of size $(2\pi)^3$ with $N^3$ grid points. While the standard resolution is $k_{max}\eta$ takes values between 
1 and 2, results are presented from $k_{max}\eta\approx 1.5$ to $11$, when available, which allows us to 
assess the effect of insufficient resolution. The Table in Fig. \ref{tab:dns_diego} summarizes those 
parameters of the DNS databased that have been used.

\subsection{The $D_m$--moments in forced stationary isotropic turbulence}

\begin{figure}
\begin{center}
\includegraphics[width=0.28\textwidth]{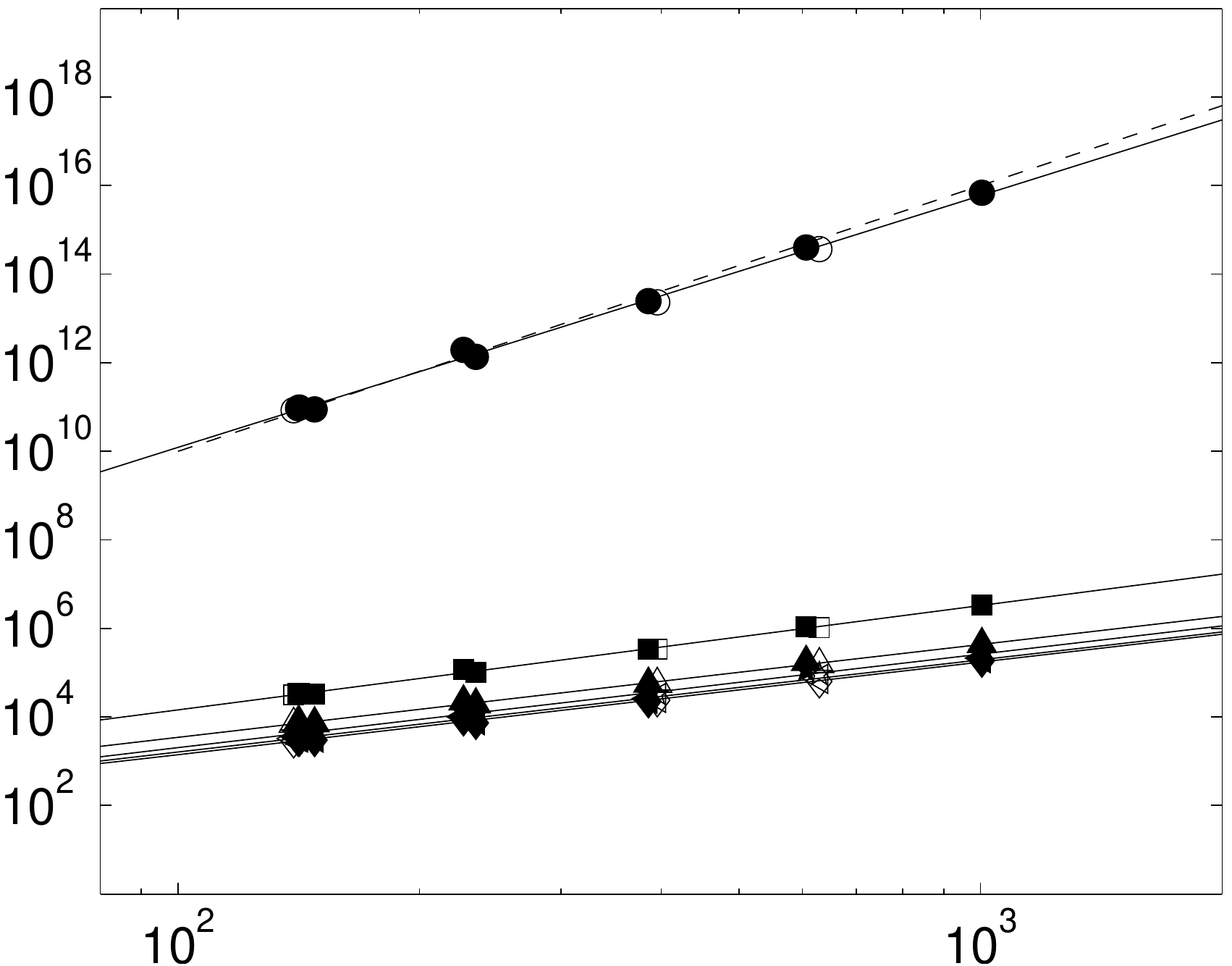}
~~
\includegraphics[width=0.28\textwidth]{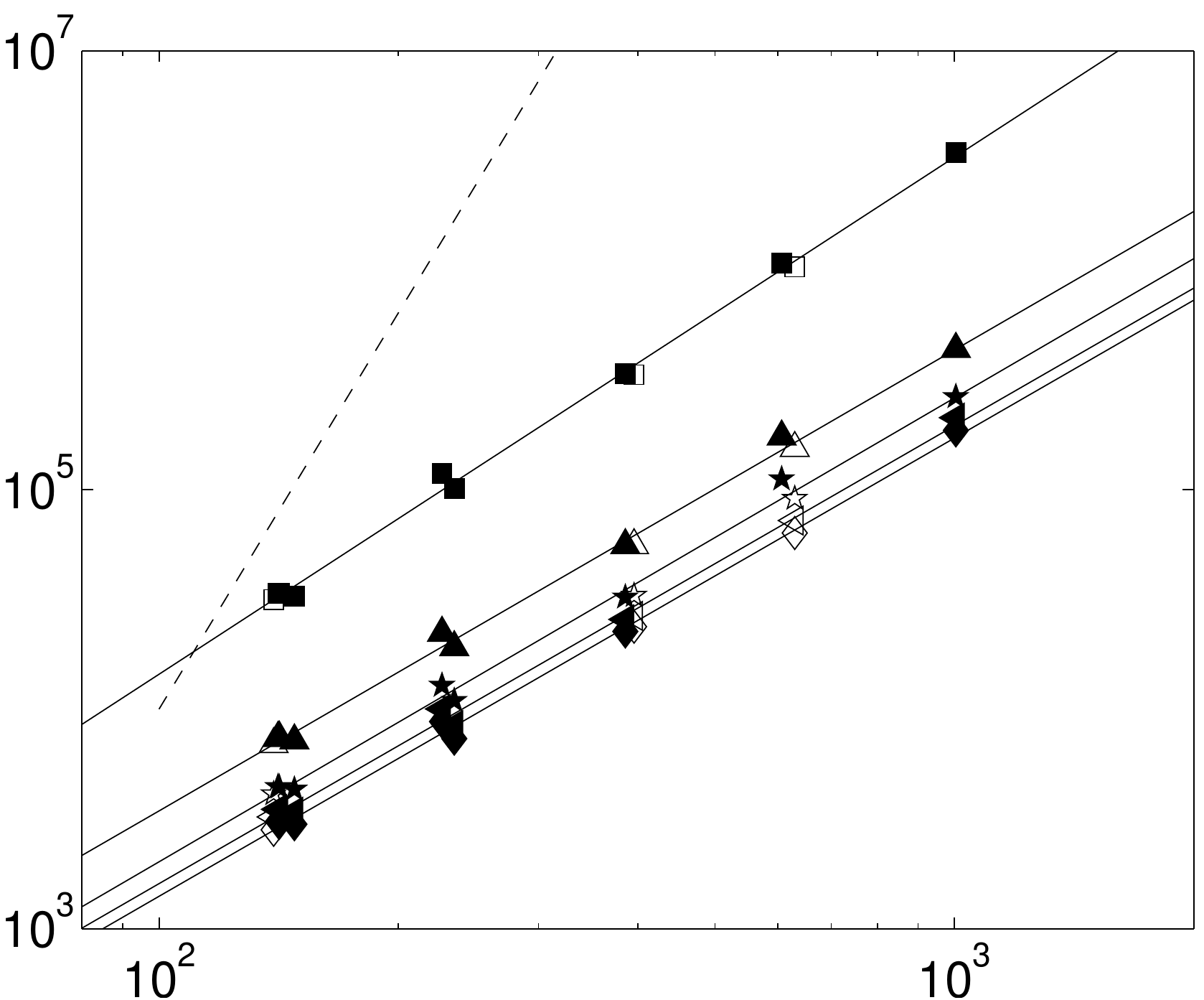}
\begin{picture}(0,0)
\put(-180,55){\vector(0,-1){35}}
\put(-178,43){\scriptsize$m$}
\put(-55,70){\vector(0,-1){35}}
\put(-55,31){\scriptsize$m$}
\put(-258,70){\scriptsize$D_m$}
\put(-200,-5){\scriptsize$Re_\lambda$}
\put(-60,-5){\scriptsize$Re_\lambda$}
\end{picture}
~~~~
\includegraphics[width=0.28\textwidth]{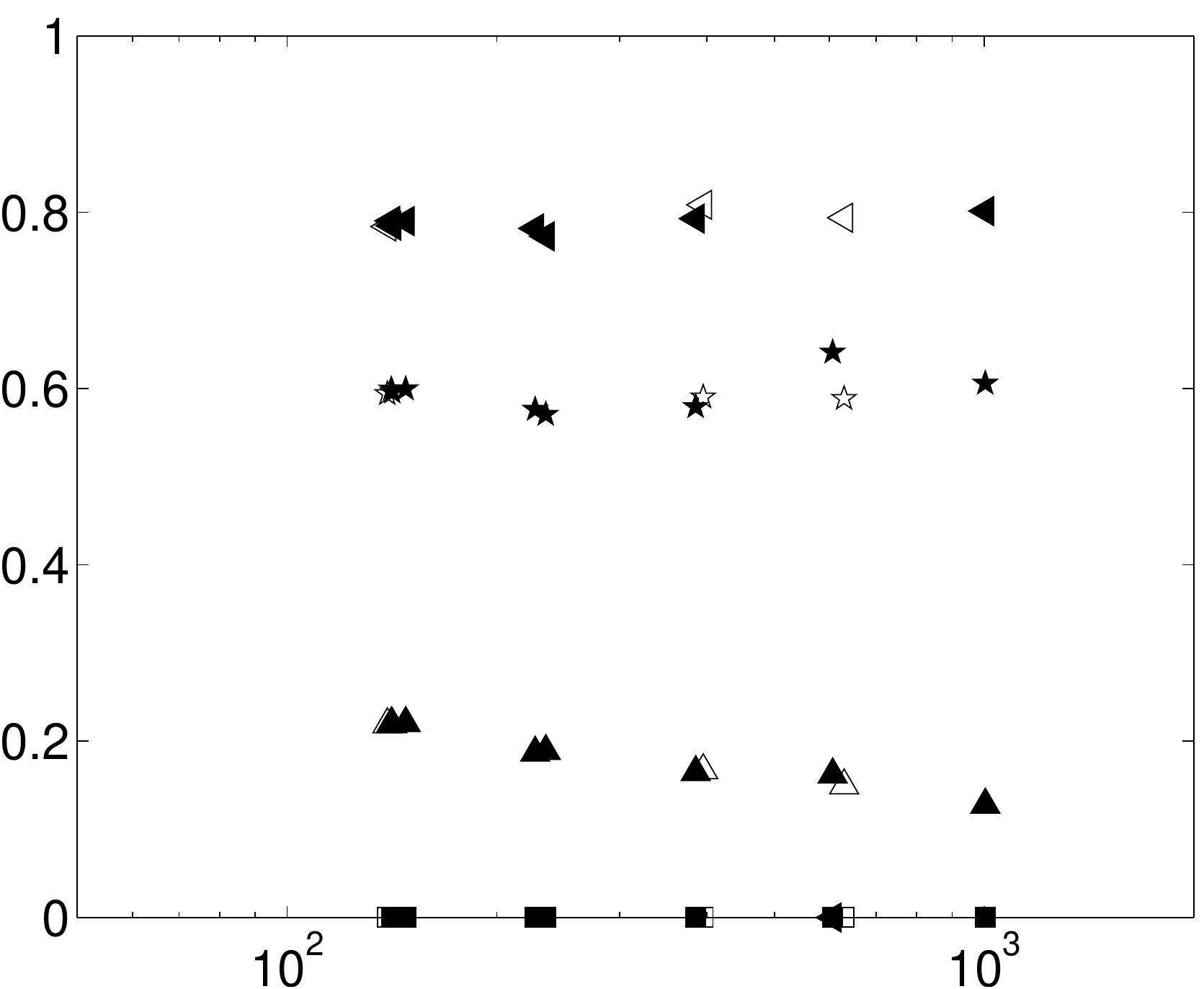}
\begin{picture}(0,0)
\put(-65,40){\vector(0,1){35}}
\put(-67,76){\scriptsize$m$}
\put(-120,45){\rotatebox{90}{\scriptsize$D_{m+1}/D_{m}$}}
\put(-70,-5){\scriptsize$Re_\lambda$}
\end{picture}
\caption{\scriptsize Scaling of moments of $D_m$ and ratios as a function of $Re_\lambda$ for 
forced stationary isotropic turbulence with resolutions up to $4096^3$.  
\textbf{First panel\,:} $D_m$ for $m=1$ to 6. \textbf{Second panel\,:}  Zoom of first 
panel to highlight the ordering of $D_m$ for $m=2$ to 6. In both parts the dashed lines 
correspond to $Re_\lambda^6$. \textbf{Third panel\,:}  Ratio of moments $D_{m+1}/D_{m}$ 
for $m=1$ (squares), $2$ (triangles), 3 (stars), and 4 (left triangles) as a function of 
$Re_\lambda$.}\label{TAMU_rat}
\end{center}
\end{figure}
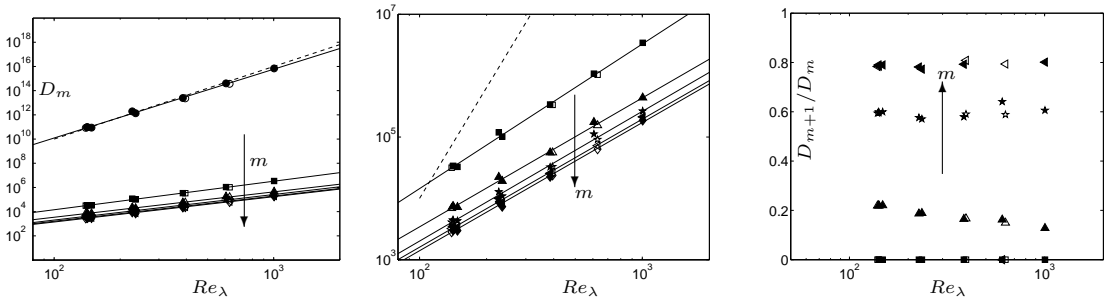
Even moments of vorticity $\Omega_{m}$ are shown in Fig.~\ref{TAMU_om}. As assured by H\"older's 
inequality it can be seen that $\Omega_{m+1}>\Omega_m$ at all Reynolds numbers. The figure also 
shows the line $\sim Re_\lambda$ (dashed), which is the result of the dissipative anomaly. This is easily 
obtained from the kinematic relation $\langle\epsilon\rangle = \nu \Omega_1^2$ associated with 
isotropic turbulence and the well-known scaling $\langle\epsilon\rangle\sim U_0^3/L$. It can then be 
shown that $\Omega_1\sim (U_0/L)Re^{1/2} \sim (U_0/L)Re_\lambda$, where the well-known result 
$Re_\lambda^2\sim Re$ has been used. The DNS data in Fig.~\ref{TAMU_om} agree 
with this scaling. As mentioned above, some resolution effects can be expected especially for high 
orders. Where data at nominally the same Reynolds number 
but different resolution is available, moments tend to be higher for higher values of $k_{max}\eta$ 
(Donzis \etal~2008). This is clearer at higher Reynolds number ($Re_\lambda\approx 650$ where 
two resolutions are available). Ratios of moments, however, are only weakly affected 
by resolution, which is also consistent with more recent results (Donzis \etal~2012, Yeung \etal~2012). 
In Fig.~\ref{TAMU_rat} the moments $D_m$ are shown as a function of $Re_\lambda$. For $m=1$, 
one can again resort to using the dissipative anomaly with the definition $D_1 = \left(\varpi_{0}^{-1}
\Omega_{1}\right)^{2}$. The result is 
\bel{Rl2}
D_1 = (L^2\sqrt{\langle\epsilon\rangle}/\nu^{3/2})^2 \sim Re^3 \sim Re_\lambda^{6}
\ee
which is seen in Fig.~\ref{TAMU_rat} (first panel). To see further details of higher order moments the 
second panel in Fig. \ref{TAMU_rat} does not include $D_1$. As in \S\ref{RMK} and \S\ref{RP}, the data 
clearly shows the ordering $D_{m+1} < D_{m}$.  The insensitive nature of moments to the type of forcing 
and the much weaker effect of resolution compared with $\Omega_m$ in Fig.~\ref{TAMU_om} is also noted.
The data also suggest that the ratio between successive moments decreases with $m$, which is consistent 
with the 
asymptotic behaviour of equation (\ref{3sim3}). This is more clearly seen in Fig.~\ref{TAMU_rat} (third panel) 
where the ratio of successive moments $D_{m+1}/D_{m}$ is plotted for different values of $m$. Consistent 
with an ordering $D_{m+1}<D_m$, the ratio is always less than unity. As $m$ increases, however, this ratio 
becomes increasingly closer to unity in agreement with Eq.~\ref{3sim3}. It is also interesting that these ratios 
appear to be independent of Reynolds numbers which suggest a regime I ordering with clustering of moments 
at high $m$ also in the high-$Re_{\lambda}$ limit. Resolution effects, while weak, can still be seen upon careful 
examination of the data, especially at high orders. However,  for a given simulation, the ordering of regime I is 
unchanged with resolution.

\section{Concluding remarks\,: the depletion of nonlinearity}\label{con}

The recent introduction of the $D_{m}$-vorticity-moment-scaling (Gibbon 2011,\,2012a,b), motivated by 
the time average (\ref{Dmav}), has suggested that they should be calculated through different numerical 
simulations.  All four dats sets unexpectedly show that the $D_m$ obey the ordering of regime I, namely 
$D_{m+1} < D_{m}$. This leads to the squeezing effect of  (\ref{3sim3}) taking place such that $\Omega_{m+1}/
\Omega_{m}\searrow 1$ and $D_{m+1}/D_{m}\nearrow 1$ as $m$ increases, which has an effect on the 
shapes of the PDF-tails, as remarked in \S\ref{sum}. The ordering in the $D_{m}$ is strict although for $m 
\geq 3,\,4$ the plots almost touch and replicate each other in shape as in Figs. \ref{Kerr}, \ref{Bang} and 
\ref{TAMU_rat} even during intense events.  It might be asked whether this is a viscous effect, or a strictly 
nonlinear effect, or the result of some surprising symbiosis between the two?  Using a variation of the 
anti-parallel initial condition used in \S\ref{RMK}, new Euler calculations have repeated this observed ordering 
(Kerr 2012c), which implicates the nonlinear terms as the source. However, there is no evidence from Navier-Stokes 
analysis that such an ordering should hold, although no results exist that suggest it cannot. It is, of course, 
possible that a cross-over could occur between regimes I and II at Reynolds numbers higher than have been 
achieved in this work. 

\begin{figure}
\includegraphics[scale=.12]{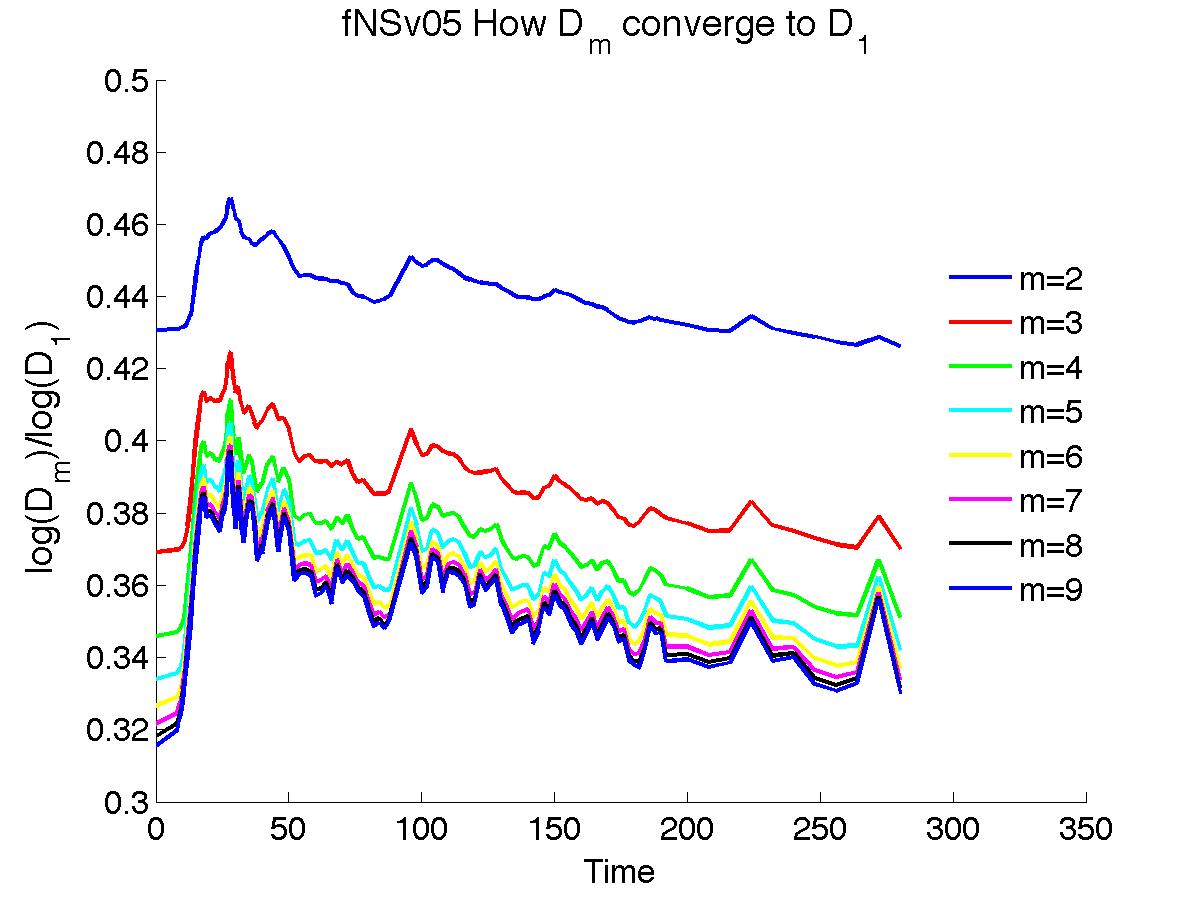} ~~~
\includegraphics[scale=.22]{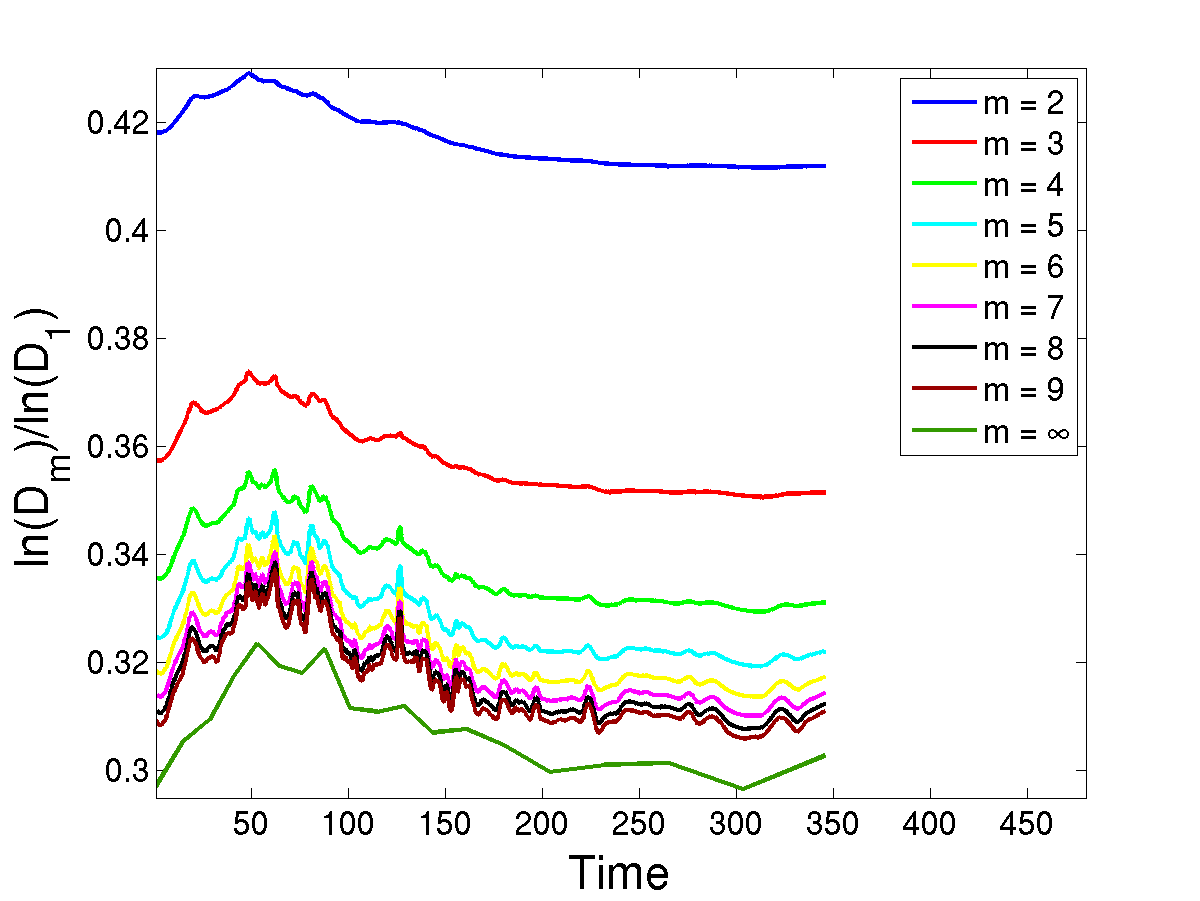}~~~
\includegraphics[scale=.22]{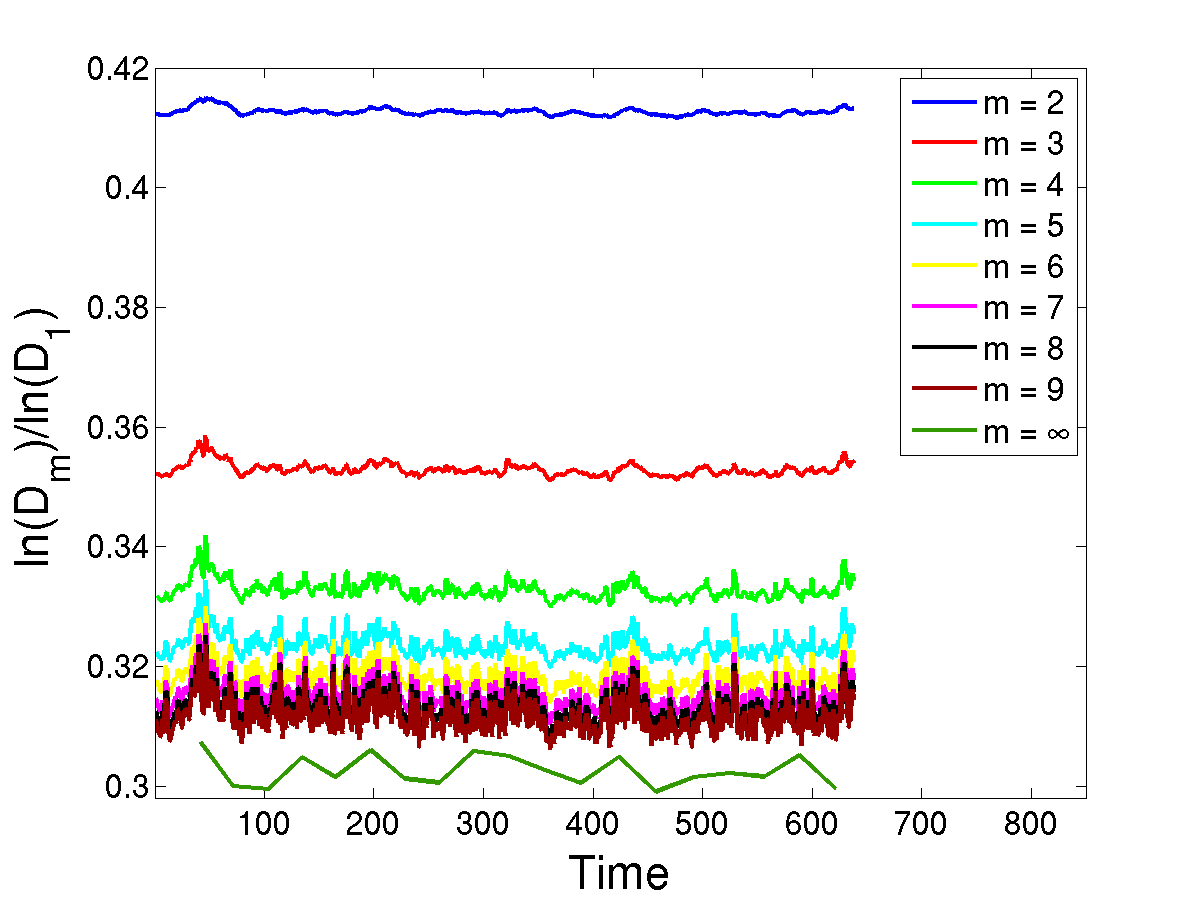}
\caption{\scriptsize Plots of $a_{m}$ for the three simulations in respectively \S\ref{RMK} and \S\ref{RP} in which 
$a_{m} < \shalf$.}\label{amfig}
\end{figure}

Significantly $D_{1}$ sits well above the other $D_{m}$ and does not appear to converge with them during the 
most intense periods\,: in Figs. \ref{Kerr}, \ref{Bang} and \ref{TAMU_rat} $D_{m}$ lies on a log-scale with 
$D_{1}$ omitted. We are therefore justified in writing
\bel{am1}
\ln D_{m} \lesssim a_{m} \ln D_{1}\qquad\Rightarrow\qquad D_{m} \lesssim D_{1}^{a_{m}}\,.
\ee
Plots of $a_{m}$ for the first and second pair of simulations are shown in Fig. \ref{amfig}. Assuming a solution 
exists, the $D_{m}$ have been shown to obey (see Gibbon 2012a) 
\beq{Dmdot1}
\dot{D}_{m} \leq D_{m}^{3}\left\{-\varpi_{1,m}\left(\frac{D_{m+1}}{D_{m}}\right)^{\frac{2}{3} m(4m+1)} + \varpi_{2,m}\right\}\,,
\eeq
where the $c_{n,m}$ within $\varpi_{1,m} = \varpi_{0}
\alpha_{m}c_{1,m}^{-1}$ and $\varpi_{2,m} = \varpi_{0}\alpha_{m}c_{2,m}$ are algebraically increasing with 
$m$. By dropping the negative term on the right hand side of (\ref{Dmdot1}), and replacing the $D_{m}^{3}$-term 
with $D_{m}D_{1}^{2a_{m}}$ justified by (\ref{am1}), a time integration produces
\bel{am2}
D_{m}(t) \leq c_{m} \exp \int_{0}^{t} D_{1}^{2a_{m}}\,d\tau \leq c_{m} 
\exp \left\{t^{1-2a_{m}}\left(\int_{0}^{t} D_{1}\,d\tau\right)^{2a_{m}}\right\}\qquad 2a_{m}\leq 1\,.
\ee
Fig. \ref{amfig} shows that while there is a weak dependence of $a_{m}$ on both $m$ and $t$, it nevertheless satisfies 
$2a_{m} < 1$ in all cases. Because Leray's energy inequality insists  that $\int_{0}^{t} D_{1}\,d\tau <\infty$ it is 
clear that the right hand side of (\ref{am2}) is finite\,: \textit{any} finite $D_{m}$ is sufficient for Navier-Stokes regularity. 
This regularization can thus be traced to the \textit{depletion of nonlinearity} in (\ref{am1}) in regime I. 
Although regime II has not been observed, (\ref{Dmdot1}) shows that it is associated with time-decay of the $D_{m}$. 
Specifically, if $D_{m+1}/D_{m} \geq \left[c_{1,m}c_{2,m}\right]^{3/2m(4m+1)}$ then $\dot{D}_{m} < 0$  where 
$\left[c_{1,m}c_{2,m}\right]^{3/2m(4m+1)} \searrow 1$ for large $m$.   

\par\vspace{3mm}\noindent
\textbf{Acknowledgements\,:} DD acknowledges the computing resources provided by the NSF-supported XSEDE and 
DOE INCITE programs under whose auspices some of these 
calculations were performed. RP and DV are members of the International Collaboration for Turbulence 
Research (ICTR). They acknowledge support from the ``Indo-French Center for Applied Mathematics'', 
UMI IFCAM -- Bangalore and, with RMK, the EU COST Action program MP0806 ``Particles in Turbulence''. 
AG and RP thank DST, CSIR and UGC (India) and the SERC (IISC) for computational resources. JDG and 
RMK thank the Isaac Newton Institute, Cambridge, on whose programme \textit{Topological dynamics in 
the Physical and Biological sciences (2012)} part of this work was carried out. 


\end{document}